\documentclass[12pt,a4paper]{article}
\pdfoutput=1
\usepackage{jheppub,amsmath,amssymb,color,youngtab,appendix}
\usepackage[nottoc]{tocbibind}
\newcommand{\bea}{\begin{eqnarray}}
\newcommand{\eea}{\end{eqnarray}}
\newcommand{\Tr}{{\rm Tr}}
\newcommand{\cH}{{\cal H}}
\newcommand{\cA}{{\cal A}}

\newcommand{\RD}[1]{\textcolor{black}{ #1}}

\usepackage{mathtools,stmaryrd,bm}
\makeatletter
\newcommand{\oset}[3][-0.25ex]{%
  \mathrel{\mathop{#3}\limits^{
    \vbox to#1{\kern-2\ex@
    \hbox{$\scriptstyle#2$}\vss}}}}
\makeatother

\makeatletter
\newcommand*{\transpose}{%
	{\mathpalette\@transpose{}}%
}
\newcommand*{\@transpose}[2]{%
	\raisebox{\depth}{$\m@th#1\intercal$}%
}
\makeatother
\begin{document}
\title{Finite $N$ Hilbert Spaces of Bilocal Holography}
\author[a,b]{Robert de Mello Koch,}
\author[c,d]{Antal Jevicki,}
\author[e,f,g]{ and Junggi Yoon,}
\affiliation[a]{School of Science, Huzhou Normal University, Huzhou 313000, China}
\affiliation[b]{Mandelstam Institute for Theoretical Physics, School of Physics, University of the Witwatersrand, Private Bag 3, Wits 2050, South Africa}
\affiliation[c]{Department of Physics, Brown University,
182 Hope Street, Providence, RI 02912, United States}
\affiliation[d]{Brown Center for Theoretical Physics and Innovation, Brown University,
340 Brook Street, Providence, RI 02912, United States}
\affiliation[e]{Department of Physics, College of Science, Kyung Hee University, Seoul 02447, Republic of Korea}
\affiliation[f]{Research Institute for Basic Sciences, Kyung Hee University, Seoul 02447, Republic of Korea}
\affiliation[g]{International Research Center for Quantum Matter, Kyung Hee University, Seoul 02447, Republic of Korea}
\date{January 2026}
\abstract{For vector/AdS and dS holography we establish the structure of the emergent Hilbert space. This is done through implementation of finite $N$ trace relations on the infinite collective space. For fermionic theories a finite Hilbert space is established, while for bosonic theories a space  of freely acting primaries multiplied by a finite set of secondaries emerges. The Hilbert space of states obey finite $N$ cut off bounds, implying finiteness of traces and entropy.}

\maketitle
\section{Introduction}

Generally, collective fields, given by the set of (single trace) invariants, offer the most systematic transition to space-time fields in holography. This is true at a dynamical level, with $1/N$ appearing as a coupling constant of the non-linear collective field theory built on an infinite overcomplete set of invariants. Even though this representation of the large $N$ theory appears to be overcomplete, in a perturbative expansion in $1/N$ it has given useful, and satisfactory results. 

At finite $N$, one still has the question of non-perturbative completeness of the theory. For a theory with $K$ degrees of freedom with $K > N$ as a rule one has finite $N$ trace relations, whose implementation is of central importance. There were indications~\cite{Jevicki:1991yi} that the theory  nevertheless applies  generally, the fundamental reason being that the finite $N$ trace relations appear to be consistent with the collective representation, in particular they  appear as null-relations preserved by the infinite space Hamiltonian. Recently~\cite{deMelloKoch:2025ngs,deMelloKoch:2025rkw,deMelloKoch:2025eqt} we have been conducting a systematic study of the space of invariants at finite $N$, and implementation of finite trace relations. The algebraic structure that emerges is succinctly given by the Hironaka decomposition, where the reduced set of invariants consists of invariants of the first kind (primary) which are freely acting and secondary invariants ,whose number is shown to grow exponentially (in powers of N). These secondary states, could be termed ``miraculous" as they surpass the finite $N$ cut off bounds. The fact that this algebraic structure can be implemented on the infinite collective space stems from the mathematical theory of Procesi~\cite{Pr}. The above algebraic reduction was seen to lead to reduction of the Hilbert space and it was most completely established on the example of theories enjoying an $S_N$ gauge invariance~\cite{deMelloKoch:2025eqt} where a $q$-deformed Hilbert space was established.
 
Vector models offer a very tractable case of AdS/dS holography~\cite {Klebanov:2002ja,Giombi:2012ms}, with the bi-local collective ~\cite{Das:2003vw,deMelloKoch:2010wdf,Das:2012dt,deMelloKoch:2018ivk,deMelloKoch:2023ngh} representation seen to provide a complete re-construction scheme,  in which not only tree but also the AdS loop expansion~\cite{deMelloKoch:2018ivk} is seen to emerge. The validity of the $1/N$ expansion stems from the exact, (finite $N$) Jacobian obtained originally in the formulation of collective theory~\cite{Jevicki:1979mb}. Indications were given that this finite $N$ measure also assures correctness at the non-perturbative level, and of the Hilbert space. This was considered in~\cite{Das:2012dt} through the geometric (K\"ahler quantization) scheme of~\cite{Berezin:1975} where the $Sp(2N)$/ de Sitter Hilbert space was fully given.

The structure of Hilbert space of a theory is closely related to its finite temperature properties. In the case of vector/AdS/dS holography this is even more pertinent. The finite temperature CFT with dual AdS  black holes, one generally sees a confinement /deconfinement phase transition appearing both in matrix~\cite{Sundborg:1999ue,Aharony:2005bq,Aharony:2003sx} and vector models~\cite{Shenker:2011zf,Amado:2016pgy}), where the importance of finite $N$ relations for  the transition is first emphasized in~\cite {Shenker:2011zf}.

Finite-$N$ effects also play a central role in the description of black hole microstates, particularly in the discovery of the fortuity mechanism~\cite{Chang:2022mjp,Choi:2022caq,Choi:2023znd,Chang:2023zqk,Choi:2023vdm,Chang:2024zqi,Kim:2025vup}: the microstates of $\frac{1}{16}$-BPS black holes fall into two classes, monotone and fortuitous. Monotone states remain BPS at all $N$, while fortuitous states lose their BPS nature above a critical $N$. In this there is definitely a similarity with the states that we  have established, and hopefully future work will illuminate the connection.

In the present work we present a comprehensive discussion of finite $N$ reduction of Hilbert space in bi-local collective theories. Full analysis  of trace relations (following our matrix and $S_N$  studies) will show consistency with the overcomplete collective and geometric quantization scheme\cite{Das:2012dt} for defining the Hilbert space of de Sitter/Sp($2N$) holography. We will perform a systematic discussion  both in bosonic and fermionic theories, with emphasis on the distinct structure of emergent Hilbert spaces in the two cases. We will then consider collective fields at finite temperature and show that the right state content is present. In particular a singlet space trace and partition function are evaluated.

\RD{A comment on terminology is in order. Throughout this paper, by ``quantization'' we mean the quantization of the \emph{finite-$K$ regulated singlet sector} in bilocal variables. The resulting Hilbert space should therefore be viewed as the regulated singlet Hilbert space written in a bilocal polarization. We are not claiming here a complete construction of the full continuum finite-$N$ field-theory Hilbert space before the regulator is removed; rather, our aim is to establish the exact finite-$N$ structure at fixed regulator and to compare it with the counting of gauge-invariant states.}

The paper is organized as follows: after reviewing the collective representation in Section \ref{CollectiveRepres}, we present a general discussion of trace relations in Section \ref{TraceRelations}, with particular emphasis on how they truncate the Hilbert space and how this truncation is structurally consistent with the overcomplete collective field theory description. Section \ref{HilbertSpace} demonstrates that the number of independent invariants, counted using the Molien-Weyl formula, exactly matches the dimension of the Hilbert space, obtained from bi-local quantization. With the Hilbert space structure in hand, traces over the Hilbert space are well defined. As an application, in Section \ref{TracePF} we compute the partition function in an illustrative example and obtain a physically sensible result. The Appendices collect various technical results.

\section{Collective Representations}\label{CollectiveRepres}

\RD{In this section we consider real $O(N)$ vector models and their singlet bilocal description. We use $O(N)$ notation throughout. When discussing the bilinear singlet sector, the distinction between $O(N)$ and $SO(N)$ is not material for the present analysis.} Collective field theory provides a description of the singlet sector, by expressing the dynamics in terms of invariant bilocal collective fields, given by
\bea
\Phi(x_1^\mu,x_2^\mu)&=&\sum_{i=1}^N\phi^i(t_1,\vec{x}_1)\phi^i(t_2,\vec{x}_2)
\eea
Even when describing fermionic vector models, the collective field is Grassmann even since it is constructed from two Grassmann odd fields $\phi^i_a$. We have in mind that the theory is defined on a spatial lattice so that the positions $\vec{x}_1$ and $\vec{x}_2$ range over a total of $K$ sites. Notice that the collective fields are bilocal both in time and space. \RD{At this stage the bilocal field is a spacetime bilocal. However, when we pass below to the canonical Hamiltonian description, we work at equal time. From that point onward $K$ counts only the regulated \emph{spatial} degrees of freedom (equivalently, spatial lattice sites or one-particle spatial modes), and time is treated separately.} In a path integral description, the change to invariant variables is accompanied by a Jacobian, producing a highly non-linear collective action
\bea
Z &=&\int d\Phi\,\ J\,e^{-S}\,\,=\,\, \int  d\Phi e^{-\,S\,+\,\log J}\,\,\equiv\,\, \int d\Phi e^{-S_{\rm coll}[\Phi]}
\eea
where, for bosonic fields we have
\bea
J_{\rm bos} &=& (\det(\Phi))^{-(K+1)+\frac{N}{2}}  
\eea
and for fermionic fields we have
\bea
J_{\rm fer} &=& (\det(\Phi))^{-(K-1)-\frac{N}{2}},
\eea
results that are found by direct evaluation, or through a consistency condition. For large $N$ expansion one uses the collective action 
while $(\det(\Phi))^{-(K\pm 1)}$ provides the integration measure (and counterterms) in the $1/N$ expansion.

 It is significant that the collective description also has, a canonical, Hamiltonian form which involves equal time invariants. \RD{The Hamiltonian below is not derived from the general spacetime bilocal action $S_{\rm coll}[\Phi]$. Rather, it is the canonical equal-time collective Hamiltonian obtained by rewriting the underlying oscillator Hamiltonian in terms of the equal-time invariants $\sigma (t,a,b)=\sum_{i=1}^N\phi_i (t,a) \phi_i(t,b)$, following the standard collective-field construction~\cite{Jevicki:1979mb}. The canonical Hilbert-space discussion is carried out at fixed time and with $K$ interpreted as the number of regulated spatial modes.} For an $N$-component real scalar field theory, one has
\bea
H_{\rm coll}&=&-\frac12\sum_{b,c,d=1}^K\sigma(c,d)\Big[
\frac{\partial}{\partial\sigma(b,d)}\frac{\partial}{\partial\sigma(b,c)}+\frac{\partial}{\partial\sigma(b,c)}\frac{\partial}{\partial\sigma(d,b)}+\frac{\partial}{\partial\sigma(c,b)}\frac{\partial}{\partial\sigma(b,d)}\cr\cr
&&\qquad\qquad +\frac{\partial}{\partial\sigma(d,b)}\frac{\partial}{\partial\sigma(c,b)}\Big]-N\sum_{a=1}^K\frac{\partial}{\partial\sigma(a,a)}+\frac12\sum_{a=1}^K\sigma(a,a)\label{Hcoll}
\eea
where the derivative acts as
\bea
\frac{\partial}{\partial\sigma(a,b)}\sigma(c,d)&=&\delta^{ac}\delta^{bd}\,.
\eea
\RD{The change of variables induces a non-trivial measure~\cite{Jevicki:1979mb}. The corresponding Hilbert-space scalar product is therefore
\bea
(F_1,F_2)=\int d\Phi\,F_1(\Phi)F_2(\Phi)\,\mu(K,N),\qquad\qquad\mu(K,N)=J(\Phi)\, .
\eea
The Hamiltonian \eqref{Hcoll} is not manifestly Hermitian, reflecting the non-trivial measure. After a similarity transformation which trivializes the measure, there results a manifestly Hermitian Hamiltonian~\cite{Jevicki:1979mb} applicable to $1/N$ expansion.}

For discussions of Hilbert space it is often useful to work in the creation-annihilation Fock space basis. It is in this basis that the Hilbert space of $Sp(N)$ de Sitter holography was first discussed~\cite{Das:2012dt}. We summarize the respective representations below.

\subsection{Bosons}
We will generally (for counting purposes) label the continuous space variable $x$ by a discrete index $k$ with $k=0,1,2,\cdots, K$  where $K$ is a momentum cut-off. For a field theory situation this cut-off is certainly larger then $N$, so $N<K$ is the range of relevance for QFT and holographic duality. We have the basic $N$-component set of creation-annihilation operators \RD{($i,j=1,…,N$)}
\bea
[\hat{a}_i(k),\hat{a}_j(l)^\dagger]&=&\delta_{ij}\delta_{kl}\qquad
[\hat{a}_i(k),\hat{a}_j(l)]\,\,=\,\,0\qquad
[\hat{a}_i(k)^\dagger,\hat{a}_j(l)^\dagger]\,\,=\,\,0
\eea
and then the $O(N)$ invariant canonical pairs:
\bea
A^\dagger (k,l)&=&\hat{a}^\dagger(k)\cdot\hat{a}^\dagger(l)\cr\cr
A(k,l)&=&\hat{a}(k)\cdot\hat{a}(l)\cr\cr
B(k,l)&=&{1\over 2} \big(\hat{a}(k)^\dagger \cdot\hat{a}(l) +  \hat{a}(l)\cdot \hat{a}(k)^\dagger\big)
\eea
Any $O(N)$ invariant operator can be expressed in terms of the basic set of (single trace) operators. We will use the following coherent state representation
\bea
\hat{a}_i(k)^\dagger\to a_i(k),\qquad\qquad \hat{a}_i(k)\to {\partial\over\partial a_i(k)}.
\eea
Then, introduce the field $\alpha(k,l)$ and its conjugate momentum 
\bea
\alpha(k,l)\equiv\sum_{i=1}^N a_i(k)a_i(l),\qquad\Pi(k,l)={\partial\over\partial\alpha(k,l)}.
\eea
Through a change of variable we obtain the collective representations
\bea
A^\dagger(k,l)&\to&\alpha(k,l),\cr\cr 
B(k,l)&\to& \sum_{q}\alpha(k,q) \Pi(q,l)+\alpha(k,l)\Pi(l,l)+ {N\over 2} \delta_{k,l}\cr\cr
A(k,l)&\to&(\Pi\alpha\Pi)(k,l)+ \Pi(k,k)(\alpha \Pi)(k,l)+ (\Pi\alpha )(k,l)\Pi(l,l)+\Pi(k,k)\alpha(k,l)\Pi(l,l)\cr
&&+(N-K-1)\Pi(k,l)+(N-K-1)\delta_{k,l}\Pi(l,l).
\eea
These operators close the $Sp(2K)$ Lie algebra, furthermore one has
\bea
\big[A^\dagger A  - B^2\big](k,l)&=&-(K+1)\big[(\alpha \Pi)(k,l)+ \alpha(k,l)\Pi(l,l)\big]-{N^2\over 4}\delta_{k,l},
\eea
implying a quadratic constraint
\bea
A^\dagger A-B^2+(1+K)B&=&{N\over 4}\big(2(K+1)-N\big)I.\label{opconstraint}
\eea
The measure in the $\alpha(k,l)$ representation is
\bea
\mu^{\rm bos}(K,N)=\det (1-Z\bar{Z})^{-(K+1)+\frac{N}{2}}
\eea
where the matrix $Z_{kl}=\alpha(k,l)$ and $\bar{Z}$ is the Hermitian conjugate of $Z$. \RD{The holomorphic matrix $Z$ is a coordinate on the compact K\"ahler phase space that arises after solving the quadratic constraint and choosing a holomorphic polarization. Accordingly, the Jacobian factors written in terms of $\Phi$ and the Berezin measure written in terms of $Z$ belong to different coordinate descriptions and should not be identified by a simple relation such as $\Phi=1-Z\bar Z$.}

\subsection{Fermions}
Similarly for the $O(N)$ symmetric system of fermions
\bea
\{\hat{b}_i(k),\hat{b}^\dagger_j(l)\}&=&\delta_{ij}\delta_{kl}\qquad
\{\hat{b}_i(k)^\dagger,\hat{b}_j(l)^\dagger\}\,\,=\,\,0\qquad
\{\hat{b}_i(k),\hat{b}_j(l)\}\,\,=\,\,0
\eea
we have the following $O(N)$ invariant pairs
\bea
A(p,q)&=& b_i(p)b_i(q),\cr\cr
A^\dag(p,q)&=& b^\dag_i(p)b^\dag_i(q),\cr\cr
B(p,q)&=& {1\over 2} \big(b^\dag_i(p)b_i(q)- b_i(q) b^\dag_i(p)\big)\,=\, {1\over 2} b^\dag_i(p) b_i(q) -{N\over 2} \delta_{p,q}
\eea
and the collective representation
\bea
\hat{b}^\dagger_i(k)\to b_i(k),\qquad\qquad\hat{b}_i(k)\to
\frac{\partial}{\partial b_i(k)}.
\eea
Introducing the field $\beta(k,l)$ and its conjugate momentum
\bea
\beta(k,l)\equiv\sum_{i=1}^Nb_i(k)b_i(l),\qquad\Pi(k,l)=\frac{\partial}{\partial\beta(k,l)}.
\eea
we have
\bea
A(k,l)&=&-(N + K -1)\Pi(k,l) - (\Pi\beta\Pi)(k,l),\cr\cr
A^\dagger(k,l)&=&\beta(k,l),\cr\cr B(k,l)&=&- \beta\Pi(k,l)-{N\over 2}\delta_{kl} .\label{fermioncollectiverep}
\eea
These operators close the $SO(2K)$ algebra, and also
\begin{equation}
    A^\dagger A + B^2 + (1 - K )B = {N\over 4}\big(N+2(K-1)\big).\label{feropconstraint}
\end{equation}
The measure in the $\beta(k,l)$ representation is
\bea
\mu^{\rm fer}(K,N)=\det (1+Z\bar{Z})^{-(K-1)-\frac{N}{2}}
\eea
where the matrix $Z_{kl}=\beta(k,l)$ and $\bar{Z}$ is the Hermitian conjugate of $Z$.

\subsection{Phase Space}
The above, non-linear relations, in particular \eqref{opconstraint} for bosons and \eqref{feropconstraint} for fermions considered in the large $N$ limit,  represent a (non-linear) phase space of these collective theories. At classical level (considering $\hbar$ as $1/N$) this was realized by 
Berezin~\cite{Berezin:1978sn}. Here (and in an accompanying publication~\cite{Us2} these will be considered at full quantum (finite $N$ level) providing an operator, group theoretic basis for the finite $N$ Hilbert space.

\section{Trace Relations and Reduction of Hilbert Space}\label{TraceRelations}

In addition to the bi-local time action representation, collective theory also
posses a canonical equal-time Hamiltonian representation. In the field theoretic case that we are studying the spacial bi-local represents a $K\times K$ matrix with $K$ infinitely large. For bosonic $O(N)$ vector models the bilocal is a symmetric $K\times K$ matrix, while for fermions the bilocal is an antisymmetric $K\times K$ matrix. Consequently both of these collective representations are overcomplete, since at large but finite $N$ one has $NK< K^2$. Nevertheless the collective representation has systematically given finite results, in perturbation expansion in $1/N$~\cite{deMelloKoch:1996mj} and most notably in obtaining both the low and the high temperature partition function~\cite{Das:2003vw,Jevicki:2014mfa}. For the later case one would expect that the finite $N$ trace relations ought to be imposed. This is certainly to be done at the canonical Hamiltonian level, in particular when discussing the Hilbert space of the theory. The reason for the consistency of the overcomplete collective representation and the finite $N$ trace relations is the fact that the trace constraint represent ``null states" of the collective Hamiltonian. This have been observed in number of previous theories~\cite{Jevicki:1991yi,deMelloKoch:2025rkw,deMelloKoch:2025eqt} and we will summarize it below for the case of bi-locals. Due to this a consistent reduction of the infinite, overcomplete Hilbert space follows, resulting in a finite Hilbert space which will be given in several representations in what follows. \RD{The purpose of the remainder of this section is to make this statement precise: we will show that the finite-$N$ trace relations define an ideal that is preserved by the collective Hamiltonian. This is the sense in which the overcomplete collective description remains consistent after quotienting by the trace relations.}

The simplest way to understand trace relations, in both bosonic and fermionic vector models, is to use the general trace-identity theorem of Procesi~\cite{Pr}. \RD{The key point is that, at fixed matrix size $N$, there exist \emph{universal} multilinear identities among traces of words in matrices given by 
\bea
\sum_{\sigma\in S_{N+1}}{\rm sgn}(\sigma) (W_1)_{i_1 i_{\sigma(1)}}(W_2)_{i_2 i_{\sigma(2)}}\cdots(W_{N+1})_{i_{N+1} i_{\sigma(N+1)}}=0.\label{ProcesiUniversal}
\eea
By substituting appropriate words into these identities one generates all trace relations of the corresponding (multi-)matrix model.}

As an illustration, consider a bosonic model with $N=2$. Since the fundamental index takes only two values, any complete antisymmetrization over three such indices vanishes. Concretely, for any three $2\times2$ matrices $W_1,W_2,W_3$, antisymmetrizing over the column indices in the product of diagonal matrix elements gives, for any values of $i_1,i_2,i_3$,
\bea
&&(W_1)_{i_1i_1} (W_2)_{i_2i_2}(W_3)_{i_3i_3}-(W_1)_{i_1i_2} (W_2)_{i_2i_1}(W_3)_{i_3i_3}-(W_1)_{i_1i_1} (W_2)_{i_2i_3}(W_3)_{i_3i_2}\cr\cr
&&-(W_1)_{i_1i_3} (W_2)_{i_2i_2}(W_3)_{i_3i_1}+(W_1)_{i_1i_2} (W_2)_{i_2i_3}(W_3)_{i_3i_1}+(W_1)_{i_1i_3} (W_2)_{i_2i_1}(W_3)_{i_3i_2}=0\cr
&&
\eea
Summing over all indices yields the universal multilinear trace identity
\bea
&&\Tr(W_1)\Tr(W_2)\Tr(W_3)-\Tr(W_3)\Tr(W_1W_2)-\Tr(W_1)\Tr(W_2W_3)\cr\cr
&&-\Tr(W_2)\Tr(W_1W_3)+\Tr(W_1W_2W_3)+\Tr(W_1W_3W_2)=0\label{universal}
\eea
Procesi's theorem~\cite{Pr} implies that, for $N=2$, inserting into (\ref{universal}) \emph{arbitrary} words built from the matrices of the model generates {\bf all} trace relations of the trace algebra.

To apply this to the vector model\footnote{\RD{For related discussions of finite-$N$ vector-model holography and bilocal variables, see Aharony et al. \cite{Aharony:2022feg}; our emphasis here is on the trace ideal, Hilbert-space reduction, and its relation to bilocal quantization.}}, start from vectors $a_i(k)$ with color index $i=1, 2, \cdots,N$ and flavor index \RD{$a,b=1,\dots,K$}. From these we build a family of matrices
\bea
(m_{ab})_{ij}&=&a_i(a)a_j(b)\,.
\eea
There are $K^2$ such matrices in the setup we consider. We focus on the case $K\ge N$. Choosing the $W_i$ in (\ref{universal}) to be \emph{any} words constructed from the $m_{ab}$ produces the complete set of trace relations for the vector model. \RD{Taking traces of these words leads to expressions that can be written in terms of the flavor bilinears 
\begin{equation}
\sigma(k,l)\,\,=\,\,\sum_{i=1}^N a_i(k)a_i(l)\,.
\end{equation}
The basic formula we use is
\begin{equation}
{\rm Tr}(m_{a_1b_1}m_{a_2b_2}\cdots m_{a_kb_k})=\sigma(b_k,a_1)\sigma(b_1,a_2)\sigma(b_2,a_3)\cdots\sigma(b_{k-1},a_k)
\end{equation}
which follows immediately upon writing both sides in terms of the vectors $a_i(k)$. In particular, $\Tr(m_{ab}m_{cd})=\sigma(b,c)\sigma(d,a)$.} For example, taking $W_1=m_{11}$, $W_2=m_{22}$ and $W_3=m_{33}$ gives $T_{123}=0$ where
\bea
T_{123}&=&\sigma(1,1)\sigma(2,2)\sigma(3,3)-\sigma(3,3)\sigma(1,2)\sigma(2,1)-\sigma(2,2)\sigma(1,3)\sigma(3,1)\cr\cr
&&\!\!\!-\sigma(1,1)\sigma(3,2)\sigma(2,3)+\sigma(1,2)\sigma(2,3)\sigma(3,1)
+\sigma(1,3)\sigma(3,2)\sigma(2,1)\label{examplerelation}
\eea
\RD{Equivalently, the finite-$N$ constraints are the vanishing of all $(N+1)\times (N+1)$ minors of the flavor matrix $\sigma(k,l)$
\bea
{\rm rank} (\sigma) &\le& N\label{flavorrelation}
\eea
This bound follows from writing $\sigma=A\,A^T$, where $A$ is the $K\times N$ matrix with entries $A_{k i}=a(k)_i$, so $\mathrm{rank}(\sigma)\le \mathrm{rank}\,A\le N$. The equivalence between the Procesi description of the trace constraints given in equation \eqref{ProcesiUniversal} and the flavor-space condition (\ref{flavorrelation}) can be verified in many ways; a quick diagnostic is to compare Hilbert series. We will discuss a concrete example in Appendix \ref{EquivRels} below. In particular, this discussion in the Appendix establishes that these are the complete set of conditions.}

A direct calculation shows that the trace relations are consistent with the collective Hamiltonian. Concretely, we must show that the collective Hamiltonian preserves the constraint ideal: it maps any state satisfying the trace relations to another state that also satisfies them. Equivalently, the collective Hamiltonian preserves the ideal generated by the trace relations. This ensures that it is consistent to impose the trace relations as operator equations on the collective Hilbert space.

We first specialize to the collective description of the real bosonic vector harmonic oscillator, whose partition function is transparently related to the Hilbert series of $O(N)$ invariants. We then indicate how the analysis generalizes to an arbitrary potential, and how it extends to complex models.
To keep the discussion explicit, introduce an operator related to $H_{\rm coll}$, given in \eqref{Hcoll}, by a similarity transformation,
\bea
\tilde{O}&=&e^{\frac12\sum_{e=1}^K\sigma(e,e)}\,H_{\rm coll}\,e^{-\frac12\sum_{e=1}^K\sigma(e,e)}\cr\cr
&=&-\frac12\sum_{b,c,d=1}^K\sigma(c,d)\Big[
{\partial\over\partial\sigma(b,d)}{\partial\over\partial\sigma(b,c)}+{\partial\over\partial\sigma(b,c)}{\partial\over\partial\sigma(d,b)}+{\partial\over\partial\sigma(c,b)}{\partial\over\partial\sigma(b,d)}\cr\cr
&&\quad+{\partial\over\partial\sigma(d,b)}{\partial\over\partial\sigma(c,b)}\Big]-N\sum_{a=1}^K{\partial\over\partial\sigma(a,a)}+\sum_{a,b=1}^K\sigma(a,b)\Big[{\partial\over\partial\sigma(a,b)}+{\partial\over\partial\sigma(b,a)}\Big]\cr\cr&&\qquad+\frac{NK}{2}\,.
\eea
The key property of $\tilde{O}$, which makes it particularly useful for the present exercise, is that the trace relations are eigenstates of  $\tilde{O}$, a fact that is immediate to check explicitly. The corresponding eigenvalue is determined in terms of the degree of the trace relation, computed with the assignment that $\sigma(a,b)$ has degree 1. For example, the trace relation $T_{123}$ given in \eqref{examplerelation} above obeys
\bea
\tilde{O}T_{123}&=&(d+\frac{NK}{2})T_{123}\,\,=\,\,(3+K)T_{123}
\eea
since the relation has degree $d=3$ and $N=2$. Multiplying any trace relation by an arbitrary polynomial yields another element of the trace relation ideal. In general, such a product need not be an eigenstate of $\tilde{O}$, and hence need not be an eigenstate of the collective Hamiltonian. Nevertheless, the action of the collective Hamiltonian preserves the trace ideal: it maps these polynomial multiples into the same ideal.

We now extend the illustrative example above to a general potential and establish the general statement that the collective Hamiltonian preserves the finite-$N$ trace relations. For potentials that depend only on the fields (and not on their conjugate momenta), the potential term acts multiplicatively in the coordinate representation and is therefore automatically compatible with the polynomial trace identities: multiplying any element of the trace-identity ideal by a polynomial produces another element of the same ideal. It follows that, for arbitrary field-dependent potentials, the full collective Hamiltonian maps the constraint ideal into itself. An entirely analogous argument applies to complex vector models, leading to the conclusion that their collective descriptions likewise preserve the finite-$N$ trace relations. The detailed discussion is completely parallel to the real free model given above so we will not repeat it. \RD{Thus the reduced Hilbert space obtained by imposing the trace relations is not only algebraically well defined, but also dynamically closed under collective time evolution.}

The description of trace relations in the fermionic vector model is completely parallel. The only new ingredient is that the fundamental variables are Grassmann-valued, so one must track signs when re-expressing trace identities in terms of flavor bilinears. 

For fermionic vectors $b_i(p)$, with color index $i=1,\ldots,N$ and flavor labels $p,q=1,\ldots,K$, we again package the color contractions into a family of $N\times N$ matrices
\bea
(n_{ab})_{ij} \;=\; b_i(a)\,b_j(b)\,.
\eea
Because each $n_{ab}$ is Grassmann-even, these matrices multiply as ordinary (bosonic) matrices, and Procesi's universal multilinear trace identities apply without modification at the level of the $n_{ab}$'s. As before, the resulting trace relations can be rewritten purely in terms of the flavor bilinears
\bea
\beta(k,l)=\sum_i b_i(k)b_i(l)\,.
\eea
The only subtlety is that, when translating a given trace identity into the $\beta$-language, moving fermions past one another can generate sign flips. A simple example is obtained by inserting $W_1=n_{11}$, $W_2=n_{22}$ and $W_3=n_{33}$ into the universal identity (\ref{universal}), which yields
\bea
&&\beta(1,1)\beta(2,2)\beta(3,3)+\beta(3,3)\beta(1,2)\beta(2,1)+\beta(2,2)\beta(1,3)\beta(3,1)\cr\cr
&&+\beta(1,1)\beta(3,2)\beta(2,3)+\beta(1,2)\beta(2,3)\beta(3,1)
+\beta(1,3)\beta(3,2)\beta(2,1)\,\,=\,\,0\cr
&&
\eea
Comparing to the bosonic case~\eqref{examplerelation}, certain sign flips are evident. Exactly as in the bosonic case, the full set of finite-$N$ constraints can be summarized as a rank condition on the $K\times K$ matrix $\beta$,
\bea
{\rm rank}\,\beta\le N\,.
\eea
This bound is immediate from writing $\beta=\Psi\,\Psi^\dagger$, where $\Psi$ is the $K\times N$ matrix with entries $\Psi_{k i}=b_i(k)$, so $\mathrm{rank}(\beta)\le \min(\mathrm{rank}\,\Psi,\mathrm{rank}\,\Psi^\dagger)\le N$. As for the bosons, one can verify that this flavor-space formulation is equivalent to the Procesi description (e.g. by comparing Hilbert series). 

In the next section we give the explicit construction of the reduced 
Hilbert space, after the imposition of finite $N$ trace relations. The structure found will be in agreement with  the associated partition functions or equivalently, the Hilbert series. From the Molien-Weyl formula, we have
\bea
Z_B&=&\frac{1 + \sum_i c^s_i x^i}{\prod_j (1 - x^j)^{c^m_j}},\label{ZB}
\eea
and 
\bea
Z_F&=&1 + \sum_i c_i x^i\label{ZF}
\eea
respectively, for the bosonic and fermionic models. Here $c^s_i$, $c^m_j$ and $c_i$ are all positive integers. For the bosonic Hilbert space, the structure of \eqref{ZB} has a natural invariant-theoretic interpretation. The denominator reflects the freely generated polynomial algebra of primary invariants, while the numerator records the finite module of secondary invariants in the corresponding Hironaka decomposition. In this sense the bosonic singlet Hilbert space contains both freely acting oscillator-type degrees of freedom and an additional finite-$N$ sector encoded by the secondary invariants; see~\cite{deMelloKoch:2025ngs} for a detailed discussion. In the examples studied so far, the number of such secondary states grows exponentially with $N$; here we use this language mainly as an efficient way to organize the counting formulas and their relation to the reduced Hilbert space.

For fermions, by contrast, \eqref{ZF} has no denominator, reflecting the absence of freely acting bosonic generators. The fermionic Hilbert space is therefore finite and is built entirely from the analogue of the secondary sector. In particular, although correlation functions of gauge-invariant operators in the bosonic $O(N)$ vector model are related to those of the fermionic $Sp(N)$ vector model by the analytic continuation $N\to -N$~\cite{Ramgoolam:1993hh,Anninos:2011ui}, this continuation does not extend to the non-perturbative Hilbert-space structure.

In fact, there is a more subtle duality operating between the two Hilbert spaces that holds at non-perturbative level. In this duality the bosonic and fermionic models have the same gauge symmetry (determined by $N$), but distinct flavor symmetry (determined by $K$). The mismatch between the flavor symmetries is crucial for the realization of the duality: the trace relations of the bosonic model are reproduced by both flavor and Grassman identities in the fermionic model. This is only possible if the bosonic and fermionic models have different flavor symmetry groups. In particular, the Hilbert spaces for theories with the same flavor symmetry are generally not equivalent. Details of these dualities will be reported elsewhere~\cite{Us}. 

\subsection{Hilbert space from trace relations}\label{BSection3}

Our aim in this section is to highlight situations in which the \emph{entire} Fock-space structure can be recovered purely from the trace relations, without invoking any additional dynamical input. The prototypical example we have in mind is a fermionic vector model at $N=1$ with $K=p$.  Here we will rederive the complete Fock space structure directly from the \emph{free bilinear algebra}, by quotienting by the ideal generated by the trace relations, thereby making explicit how the full state space is fixed algebraically.

We have $p$ fermionic modes with creation/annihilation operators $\{a_i,a_i^\dagger\}_{i=1}^p$ satisfying the canonical anti-commutation relations. Let $\Phi_0$ be the vacuum, $a_i\Phi_0=0$, and let $\cH$ be the Fock space spanned by
\begin{equation}
a_{i_1}^\dagger \cdots a_{i_n}^\dagger \Phi_0,\qquad 1\le i_1<\cdots<i_n\le p.
\end{equation}
We know that $\dim\cH=2^p$. There is a decomposition
\begin{equation}
\cH=\cH_{\rm even}\oplus \cH_{\mathrm{odd}}.
\end{equation}
$\cH_{\mathrm{even}}$ is spanned by states with even fermion number and this is the gauge invariant sector of the theory. \RD{For $N=1$, the gauge group is $O(1)\simeq{\mathbb Z}_2$, under which each fermionic creation operator is odd, $a_i^\dagger\mapsto -a_i^\dagger$. Hence only states with even fermion number are gauge invariant.} To describe this sector we introduce the \emph{collective bilinear invariants}
\begin{equation}
Z_{ij}=a_i^\dagger a_j^\dagger,\qquad Z_{ij}=-Z_{ji},\qquad Z_{ii}=0.
\end{equation}
These are \emph{even} operators so that acting with any polynomial in the $Z_{ij}$ on $\Phi_0$ produces an even state. The gauge invariant Fock space $\cH_{\rm even}$ is generated by acting on $\Phi_0$ with the $Z_{ij}$. We know that $\dim\cH_{\rm even}=2^{p-1}$.

The $Z_{ij}$'s are products of two fermion oscillators, so they commute
\begin{equation}
Z_{ij}Z_{kl}=Z_{kl}Z_{ij}.
\end{equation}
It is natural to model the span of states generated by the $Z_{ij}$ as a \emph{commutative} algebra on antisymmetric generators. Let $\cA_{\mathrm{free}}$ be the commutative algebra over $\mathbb C$ generated by symbols $\{Z_{ij}\}_{1\le i<j\le p}$ with $Z_{ji}=-Z_{ij}$ and $Z_{ii}=0$. Let $I$ be the ideal generated by the two families of quadratic relations
\begin{align}
\text{(R1)}\qquad & Z_{ij}Z_{il}=0 \quad\text{for all }i,j,l, \label{R1}\\
\text{(R2)}\qquad & Z_{ij}Z_{kl}+Z_{il}Z_{kj}=0 \quad\text{for all }i,j,k,l. \label{R2}
\end{align}
Define
\begin{equation}
\cA = \cA_{\mathrm{free}}/I.
\end{equation}
Our basic claim is that the quotient algebra $\cA$ is isomorphic to $\cH_{\rm even}$ which provides a demonstration of how the complete Fock space structure is recovered from the trace relations.

A comment is in order. (R1) follows because we study fermions. This is easily demonstrated explicitly:
\begin{equation}
Z_{ij}Z_{il}=(a_i^\dagger a_j^\dagger)(a_i^\dagger a_l^\dagger)= a_i^\dagger a_j^\dagger a_i^\dagger a_l^\dagger
= - a_i^\dagger a_i^\dagger a_j^\dagger a_l^\dagger=0,
\end{equation}
On the other hand, (R2) is the $N=1$ trace relation and can be derived from Procesi's universal multilinear identity described above.

To prove our basic claim, we define a linear map
\begin{equation}
\pi:\cA_{\mathrm{free}}\longrightarrow\cH_{\rm even},\qquad\pi(P)= P\bigl(a_i^\dagger a_j^\dagger\bigr)\,\Phi_0.
\end{equation}
Because (R1) and (R2) hold for the operators, $I\subset\ker\pi$, $\pi$ descends to a well-defined map
\begin{equation}
\overline\pi:\cA=\cA_{\rm free}/I\longrightarrow \cH_{\rm even}.
\end{equation}
First we want to argue that $\overline\pi$ is surjective, i.e. all of $\cH_{\rm even}$ lies in the image of $\overline\pi$. To see this is true, note that any even basis vector of $\cH$ has the form
\begin{equation}
\Phi_{i_1\cdots i_{2m}}\,\,=\,\,a_{i_1}^\dagger\cdots a_{i_{2m}}^\dagger \Phi_0\qquad (i_1<\cdots<i_{2m}).
\end{equation}
But
\begin{equation}
\Phi_{i_1\cdots i_{2m}} = \pm (a_{i_1}^\dagger a_{i_2}^\dagger)\cdots(a_{i_{2m-1}}^\dagger a_{i_{2m}}^\dagger)\Phi_0= \pm Z_{i_1 i_2}\cdots Z_{i_{2m-1} i_{2m}}\Phi_0.
\end{equation}
Thus every even basis vector lies in the image of $\pi$, so $\overline\pi$ is indeed surjective. We now need to prove injectivity, i.e. do our two families of relations generate all linear dependencies among monomials $Z_{i_1j_1}\cdots Z_{i_mj_m}\Phi_0$? A clean way to prove this is to show that our relations reduce every monomial to a \emph{unique canonical representative} labelled by an even subset $I\subset\{1,\dots,p\}$, and that these canonical representatives map to the standard basis of $\cH_{\rm even}$. Given an even subset $I=\{i_1<\cdots<i_{2m}\}$, define the canonical monomial
\begin{equation}
M_I = Z_{i_1 i_2}\,Z_{i_3 i_4}\cdots Z_{i_{2m-1} i_{2m}}\quad\in \cA.
\end{equation}
It is not difficult to argue that the $M_I$ span $\cA$, and are linearly independent. Thus, a canonical basis of $\cA$ is labelled by even subsets. 

Every monomial in $\cA$ is equal to a linear combination of $M_I$'s. Consider any monomial $Z_{i_1 j_1}\cdots Z_{i_m j_m}$. If any index is repeated the product vanishes by (R1). So nonzero monomials must have \emph{all} $2m$ indices distinct. If all indices are distinct, the monomial is (up to sign) the unique $2m$-fermion state $a_{r_1}^\dagger\cdots a_{r_{2m}}^\dagger\Phi_0$ where $\{r_a\}$ is the set of indices appearing. Different pairings of a set of indices correspond to different factorizations into $Z$'s. Relation (R2) is the elementary ``flip'' on four distinct indices that changes one pairing into another. By repeated flips any pairing can be converted into the canonical pairing $(i_1,i_2)(i_3,i_4)\cdots(i_{2m-1},i_{2m})$ after sorting the indices. This expresses any monomial as \(\pm M_I\). Consequently, the $M_I$ span $\cA$.

Next, apply $\overline\pi$ to $M_I$
\begin{equation}
\overline\pi(M_I) = (a_{i_1}^\dagger a_{i_2}^\dagger)\cdots(a_{i_{2m-1}}^\dagger a_{i_{2m}}^\dagger)\Phi_0 = \pm a_{i_1}^\dagger\cdots a_{i_{2m}}^\dagger\Phi_0.
\end{equation}
Distinct subsets $I$ give distinct basis vectors of $\cH_{\rm even}$ and are thus linearly independent. Therefore, the set $\{M_I\}$ is linearly independent in $\cA$ and hence $\dim\cA = \sum_m \binom{p}{2m}=2^{p-1}$. This completes the proof of injectivity.

To confirm our logic in some concrete examples, we have used the open source software system Macaulay2. Using Macaulay2 we can specify the free polynomial ring $R$ generated by the bilinears and the ideal $I$ generated by the relations (R1) and (R2). Macaulay2 can then be used to evaluate the Hilbert series of the quotient algebra $A=R/I$ and further, to provide an explicit basis for this quotient algebra. After choosing a value for $p$ these computations confirm that $A$ indeed realizes the Hilbert space generated by the collective bilinear. In Appendix \ref{Macaulay2} we have given examples of scripts for $p=2$ and $p=4$.

This basis, representing the $N=1$ bi-local Hilbert space, was also identified by Berezin~\cite{Berezin:1975}, in K\"ahler quantization. The corresponding quantization produces a finite-dimensional Hilbert space of holomorphic functions, admitting an orthonormal basis
\begin{equation}
Q_0(z)=1,\qquad Q_{i_1\cdots i_{2m}}(z)\ \ \ (1\le 2m\le p),
\end{equation}
where $Q_{i_1\cdots i_{2m}}(z)$ is the Pfaffian of the $2m\times 2m$ principal skew submatrix $z_{I}$ with $I=\{i_1,\cdots,i_{2m}\}$, equivalently 
$Q_I(z)^2=\det z_I$. A key identity,
\begin{equation}
\bigl[\det({\bf 1}+zz^*)\bigr]^{1/2}
=\sum_{\substack{I\subset{1,\dots,p}\ |I|\ \mathrm{even}}}|Q_I(z)|^2,
\end{equation}
implies that these $Q_I$ form an orthonormal basis with respect to Berezin’s inner product at the special value $h_0=2(p-1)$. Moreover, Berezin identifies the resulting Hilbert space with a chiral spinor representation of $O^+(2p,\mathbb R)$. We refer the reader to Section~3 of \cite{Berezin:1975} for the original presentation and to Section \ref{HilbertSpace} for further related discussion. We now zero in on the general $(N,K)$ Hilbert space characterization, which we state in group theory terms. Namely the finite $N,K$ Hilbert space can be seen as the irreducible representation with highest weight state $\{\frac{N}{2},\frac{N}{2},...,\frac{N}{2}\}$ of the $O(2K)$ group operating on the Hilbert space. We give evidence for this by comparing the dimensionality of this representation with the counting  of singlet states through the Molien-Weyl formula. First we compute the dimension of this irreducible representation using the Weyl character formula. The result is
\begin{equation}
\dim V_{\{\frac{N}{2},\frac{N}{2},...,\frac{N}{2}\}}\equiv{\rm Dim}(N,2k)=\prod_{1\le i<j\le k}\frac{N+2k-i-j}{\,2k-i-j\,}.\label{eq:dimformula}
\end{equation}
This is in complete agreement with the formula which counts invariants using the Molien-Weyl formula. The Molien-Weyl result, given in \eqref{SOPFnctn}, is
\begin{eqnarray}
Z_{O(2N)}^{K}(1)&=&\frac{1}{2^{N}\,N!}\,\prod_{j=0}^{N-1}\frac{(j+1)!\,(2j)!\,(2K+2j)!}{j!\,(K+j)!\,(K+N+j-1)!} 
\end{eqnarray}
It is simple to verify that
\begin{equation}
{\rm Dim}(2N,2k)\,\,=\,\,Z_{O(2N)}^{2k}(1)
\end{equation}
which is a convincing check. Finally, note that the representation we have identified also follows from general representation-theory argument employing the Cartan product~\cite{Eastwood}.

\section{Counting: Molien-Weyl vs Hilbert space}\label{HilbertSpace}
As used above the Molien-Weyl formula gives a useful (finite $N$ and $K$) expression for the free energy of the Gaussian bosonic or fermionic model.
With temperature set to $\infty$ (equivalently  $x=1$) one gets a count of the total number of states in Hilbert space. We will consider, from now on, the fermionic case (for which the number is finite). This exact number will be compared with the reduced singlet collective Hilbert space built from $K$ pairs of fermionic fields transforming in the fundamental and antifundamental representations of the $U(N)$ gauge group. As described above there are several equivalent representations of this Hilbert space. We find it most convenient to use the complex representation (related with the bi-local creation-annihilation) operator basis. On one hand, this is obtained by a change of variables to complex collective invariants, with an exact collective measure defining the Hilbert space. Berezin\cite{Berezin:1975,Berezin:1978sn} has, at large $N$  identified the resulting space of configurations as a K\"ahler phase space, and introduced a generalized ``geometric quantization" using $1/N$ as $\hbar$. The number of states in this Hilbert space can then be evaluated by calculating $\Tr({\bf 1 })$ with ${\bf 1}$ the identity on Hilbert space. This evaluation was already featured in \cite{Das:2012dt} emphasising agreement at large $N$. Here we perform an exact comparison for all $N$ and $K$  with the exact number of states given by the Molien-Weyl formula. Our main result is an exact agreement: the number of independent invariants matches the dimension of the Hilbert space obtained from bi-local quantization. \RD{In particular, the comparison performed in this section is at fixed regulator $K$: it concerns the regulated singlet Hilbert space.}

\subsection{Molien-Weyl counting of fermionic singlets}\label{countingfermions}

The derivation of the Molien-Weyl formula for bosonic vector models is given in \cite{deMelloKoch:2025cec}. In this section we explain the derivation for fermionic vector models. We will discuss the case of $U(N)$ gauge symmetry but the extension to other gauge groups is obvious.

Gauge-invariant states are constructed from products of fields with color indices fully contracted. The basic gauge-invariant operator is of the form $\bar{\psi}^a\cdot\psi^b$, where $a,b$ label different flavors. To track the various fields of the theory, introduce two sets of chemical potentials $\mu_{\psi^a}$ and $\mu_{\bar{\psi}^a}$. The partition function is written in terms of the variables
\bea
x^a=e^{-\beta E_{\psi^a} -\mu_{\psi^a}}\qquad
y^a=e^{-\beta E_{\bar{\psi}^a} -\mu_{\bar{\psi}^a}}
\eea
For a free theory the exact partition function is~\cite{Sundborg:1999ue,Aharony:2003sx}
\bea
Z(x^a,y^a)&=&\sum_{\{n_a,m_a\}\ge 0}\#(\{n_a,m_a\})\,\prod_{a=1}^{f}(x^a)^{n_a}(y^a)^{m_a}\,.
\eea
where $\#(\{n_a,m_a\})$ is the number of singlets that can be constructed using $n_a$ $\psi^a$ fields and $m_a$ $\bar{\psi}^a$ fields. Using characters we can write the number of singlets as an integral over $U(N)$ to obtain
\bea
Z(x^a,y^a)&=&\int_{U(N)}[DU]\,\prod_{a=1}^{f}\left(\sum_{n_a=0}^{\infty}(x^a)^{n_a}\chi_{\wedge^{n_a}F}(U)
\right)\left(\sum_{m_a=0}^{\infty}(y^a)^{m_a}\chi_{\wedge^{m_a}\bar F}(U)\right).\nonumber
\eea
$F,\bar{F}$ stand for the fundamental and anti-fundamental representations. The anti-symmetric product of these representations reflects the fact that our fields are fermionic. After making use of the identity
\bea
\sum_{n=0}^\infty x^n \chi_{\wedge^n R}(U)={\rm exp}\left(-\sum_{m=1}^\infty \frac{(-x)^m}{m}\chi_R(U^m)\right)
\eea
the partition function becomes
\bea
Z(x^a,y^a)&=&\int_{U(N)} [DU]\,\,{\rm exp}\left(-\sum_{a=1}^f \sum_{m=1}^\infty {(-x^a)^m\over m}\chi_F(U^m)-\sum_{a=1}^f \sum_{m=1}^\infty {(-y^a)^m\over m}\chi_{\bar{F}}(U^m)\right)\cr
&&
\eea
Now performing a parallel analysis to that performed in \cite{deMelloKoch:2025cec}, we rewrite this as
\bea
Z(x^a,y^a)&=&\frac{1}{N!(2\pi i)^N}\oint\prod_{l=1}^N \frac{d\varepsilon_l}{\varepsilon_l}\,
\Delta_+(\varepsilon)\,\Delta_-(\varepsilon)\,\prod_{a=1}^{f}\prod_{l=1}^{N}
(1+x^a\varepsilon_l)(1+y^a\varepsilon_l^{-1}),\label{secfinalPF}
\eea
with
\bea
\Delta_+(\varepsilon)=\prod_{1\le k<r\le N}(\varepsilon_k-\varepsilon_r),
\qquad
\Delta_-(\varepsilon)=\prod_{1\le k<r\le N}(\varepsilon_k^{-1}-\varepsilon_r^{-1}).
\eea
Each integral over the variables $\varepsilon_i$ is performed along the unit circle in the complex plane and can be evaluated using the residue theorem. After fixing a specific value of $N$ in (\ref{secfinalPF}), the resulting expression is used to evaluate the exact partition functions used in the paper. 

We now apply the Molien-Weyl formula to a model of $2K$ fermionic fields, $K$ in the fundamental representation of $U(N)$ and $K$ in the anti-fundamental of $U(N)$. This counts the number of $U(N)$ singlets we can obtain from these $2K$ fields. Using the Molien-Weyl formula discussed above, the result for the Hilbert series is
\bea
Z(x)&=&\frac{1}{N!(2\pi i)^N}\oint \prod_{j=1}^N \frac{d\varepsilon_j}{\varepsilon_j}\;\Delta_+(\varepsilon)\,\Delta_-(\varepsilon)\;
\prod_{i=1}^N (1+\varepsilon_i x)^K(1+\varepsilon_i^{-1}x)^K,\label{finaluPF}
\eea
\bea
\Delta_+(\varepsilon)&=&\prod_{1\le k<r\le N}(\varepsilon_k-\varepsilon_r),\qquad
\Delta_-(\varepsilon)\,\,=\,\,\prod_{1\le k<r\le N}(\varepsilon_k^{-1}-\varepsilon_r^{-1}).
\eea
The integral over each $\varepsilon_j$ is over a unit circle in the complex $\varepsilon_j$ plane. After fixing specific values of $N$ and $K$ in (\ref{finaluPF}), we can integrate to evaluate the exact partition function, using the residue theorem. However, it is also possible to evaluate this integral analytically when $x=1$, for any $N$ and $K$. First, note that (\ref{finaluPF}) is the Weyl integration formula for a $U(N)$ Haar integral over eigenvalues
\bea
Z(x)&=&\int_{U(N)} dU\;\det(1+xU)^K\,\det(1+xU^\dagger)^K\,,\label{eq:unitary_form}
\eea
where $dU$ is the normalized Haar measure. The number of singlets is obtained by setting $x=1$
\bea
Z(1)&=&\int_{U(N)} dU\;|\det(1+U)|^{2K}\,.\label{eq:dim_is_Z1}
\eea
We will evaluate $Z(1)$ analytically and show that it equals
\bea
Z(1)&=&\prod_{j=0}^{K-1}\frac{\Gamma(j+1)\Gamma(N+K+j+1)}{\Gamma(K+j+1)\Gamma(N+j+1)}.
\label{counting_app_goal}
\eea
in complete agreement with the result (\ref{DimensionH}) which is derived in the next section and follows as a result of K\"ahler quantization. Using $\Delta_-(\varepsilon)=\prod_{k<r}(\varepsilon_k^{-1}-\varepsilon_r^{-1})$, rewrite the Vandermonde product as
\bea
\Delta_+(\varepsilon)\,\Delta_-(\varepsilon)&=&\prod_{1\le k<r\le N}(\varepsilon_k-\varepsilon_r)(\varepsilon_k^{-1}-\varepsilon_r^{-1})\,\,=\,\,\prod_{1\le k\neq r\le N}\left(1-\frac{\varepsilon_k}{\varepsilon_r}\right)\,.
\label{eq:vand_prod_as_A}
\eea
This follows because for each unordered pair $\{k,r\}$ we have 
\bea
(\varepsilon_k-\varepsilon_r)(\varepsilon_k^{-1}-\varepsilon_r^{-1})&=&-\frac{(\varepsilon_k-\varepsilon_r)^2}{\varepsilon_k\varepsilon_r}\,\,=\,\,(1-\frac{\varepsilon_k}{\varepsilon_r})(1-\frac{\varepsilon_r}{\varepsilon_k})
\eea
Also note that
\bea
(1+\varepsilon_i)^K(1+\varepsilon_i^{-1})^K=\varepsilon_i^{-K}(1+\varepsilon_i)^{2K}.
\eea
Thus the integrand in \eqref{finaluPF} is a Laurent polynomial in each $\varepsilon_i$. Since
\bea
\frac{1}{2\pi i}\oint \frac{d\varepsilon}{\varepsilon}\,\varepsilon^m&=&\delta_{m,0}\,,
\eea
the $N$-fold contour integral simply extracts the \emph{constant term} (CT) in all variables
\bea
Z(1)&=&\frac{1}{N!}\;\mathrm{CT}_{\varepsilon_1,\dots,\varepsilon_N}
\left[\prod_{1\le k\neq r\le N}\left(1-\frac{\varepsilon_k}{\varepsilon_r}\right)
\prod_{i=1}^N (1+\varepsilon_i)^{2K}\,\varepsilon_i^{-K}\right].\label{eq:constant_term_form}
\eea
This is a standard form of a \emph{Macdonald--Morris constant term} for the $A_{N-1}$ root system~\cite{AnSelberg}. There is a known closed formula for this constant term~\cite{AnSelberg}.  In the special case we need we have
\bea
\mathrm{CT}_{\varepsilon_1,\dots,\varepsilon_N}\left[\prod_{1\le k\neq r\le N}\left(1-\frac{\varepsilon_k}{\varepsilon_r}\right)\prod_{i=1}^N (1+\varepsilon_i)^{2K}\,\varepsilon_i^{-K}\right]&=&
N!\,\prod_{m=0}^{N-1}\frac{\Gamma(m+1)\Gamma(2K+m+1)}{\Gamma(K+m+1)^2}.
\label{eq:morris_eval}\cr
&&
\eea
Inserting \eqref{eq:morris_eval} into \eqref{eq:constant_term_form} immediately gives
\bea
Z(1)&=&\prod_{m=0}^{N-1}\frac{\Gamma(m+1)\Gamma(2K+m+1)}{\Gamma(K+m+1)^2}.
\label{eq:hua_form}
\eea
To proceed, use the Barnes $G$-function, defined by $G(1)=1$ and $G(z+1)=\Gamma(z)\,G(z)$. It implies the useful product identity
\bea
\prod_{m=0}^{N-1}\Gamma(m+a)=\frac{G(N+a)}{G(a)}\,.\label{eq:Barnes_product}
\eea
Apply \eqref{eq:Barnes_product} to \eqref{eq:hua_form}, to find
\bea
Z(1)
&=&\frac{G(N+1)\,G(N+2K+1)\,G(K+1)^2}{G(2K+1)\,G(N+K+1)^2}\,.\label{eq:Z1_Barnes}
\eea
Now compute the right-hand side of \eqref{counting_app_goal} in the same way
\bea
\prod_{j=0}^{K-1}\frac{\Gamma(j+1)\Gamma(N+K+j+1)}{\Gamma(K+j+1)\Gamma(N+j+1)}
&=&\frac{\displaystyle \left(\prod_{j=0}^{K-1}\Gamma(j+1)\right)\left(\prod_{j=0}^{K-1}\Gamma(N+K+j+1)\right)}
{\displaystyle \left(\prod_{j=0}^{K-1}\Gamma(K+j+1)\right)\left(\prod_{j=0}^{K-1}\Gamma(N+j+1)\right)}\cr\cr
&=&\frac{G(N+1)\,G(N+2K+1)\,G(K+1)^2}{G(2K+1)\,G(N+K+1)^2}\,.\label{eq:rhs_Barnes}
\eea
The Barnes-$G$ expressions \eqref{eq:Z1_Barnes} and \eqref{eq:rhs_Barnes} coincide identically, proving \eqref{counting_app_goal}. 

\subsection{K\"ahler Quantization}

Bi-local fields become finite dimensional matrices of size $K$, and due to the quadratic (matrix) constraints form a compact symmetric (Kahler) space
\begin{equation}
ds^2=\text{tr}[dZ(1-\bar{Z}Z)^{-1}d\bar{Z}(1-Z\bar{Z})^{-1}]
\end{equation}
For the hermitean matrix case we have the  manifold $M_I(K,K)$ in the classification of \cite{Berezin:1975} with the K\"ahler scalar product given by
\begin{equation}
(F_1,F_2)=C(N,K)\int d\mu(\bar{Z},Z)F_1(Z)F_2(\bar{Z})\det [1+\bar{Z}Z]^{-N}
\end{equation}
where the (K\"ahler) integration measure is given by
\begin{eqnarray}
d\mu(\bar{Z},Z)=\det[1+\bar{Z}Z]^{-2K}d\bar{Z}dZ
\end{eqnarray}
The normalization constant $C(N,K)$ is found by requiring $(F_1,F_1)=1$ for $F_1=1$. Setting
\begin{equation}
a(N,K)=\frac{1}{C(N,K)}=\int d\mu(\bar{Z},Z)\det [1+\bar{Z}Z]^{-N}
\end{equation}
we need to evaluate the matrix integral (the complex Penner Model) 
\begin{equation}
a(N,K)=\frac{1}{C(N,K)}=\int \prod_{k,l=1}^K d\bar{Z}(k,l)dZ(k,l)\det [1+\bar{Z}Z]^{-2K-N}
\end{equation}
which determines $C(N,K)$. One can show that $N$ takes only integer values. Further, the compact nature of the phase space ensures that the Hilbert space is finite dimensional. The dimension of the Hilbert space is given by
\begin{equation}
\text{Tr}(I)=C(N,K)\int \prod_{k,l=1}^K d\bar{Z}(k,l)dZ(k,l)\det [1+\bar{Z}Z]^{-2K}
\end{equation}
This matrix integral integral has been performed in \cite{Das:2012dt} by mapping it to a class of integrals evaluated by Selberg \cite{Selberg:1944}. The result is
\begin{equation}
\text{Dim  $\mathcal{H}$}_B=\prod_{j=0}^{K-1}\frac{\Gamma(j+1)\Gamma(N+K+j+1)}{\Gamma(K+j+1)\Gamma(N+j+1)} \label{DimensionH}
\end{equation}

\section{Trace and Partition Function}\label{TracePF}
\begin{figure}[htbp]
  \centering
  \includegraphics[width=0.95\textwidth]{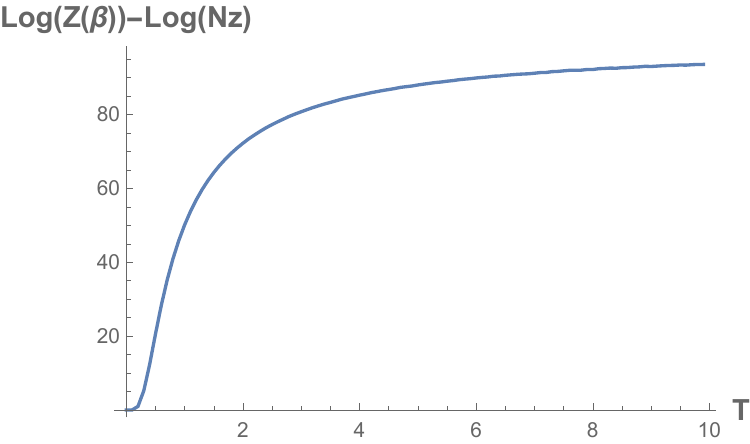}
  \caption{A plot of the logarithm of the normalized partition function, $\log\!\big(Z(\beta)/N_Z\big)$, as a function of the temperature $T$ (with $\beta \equiv 1/T$). The temperature is measured in units of $\omega^{-1}$. The example shown has $N=10$ and $K=12$.}
  \label{fig:Z}
\end{figure}

In this final section we give a specification of the trace and corresponding 
evaluation of the partition function in the reduced, singlet Hilbert space. We will detail the fermionic $U(N)$, $O(N)$ and $Sp(2N)$ invariant spaces which
represent finite Hilbert spaces of bi-local Holography. As in~\cite{Das:2012dt} we use the holomorphic (complex) representation of bi-locals
\bea
\hat{\beta}(k,l)\to Z(k,l)\qquad \hat{\beta}^\dagger(k,l)\to \bar{Z}(k,l)
\eea
with the scalar product
\bea
(F_1,F_2)&=&C(N,K)\int (d\bar{Z} dZ)\mu_{NK}(\bar{Z},Z)F_1(\bar{Z})F_2(Z)
\eea
where the measure reads
\bea
\mu_{NK}(\bar{Z,}Z)&=&\det ({\bf 1}+\bar{Z}Z)^{-2K-N}
\eea
with wave functions $F(Z)$ given as polynomials and related to states as
\bea
F(Z)&=&\langle Z|F\rangle
\eea
and
\bea
\int [dZd\bar{Z}] |Z\rangle\langle\bar{Z}|={\bf 1}
\eea
with
\bea
[dZd\bar{Z}]&=&C(N,K) (d\bar{Z} dZ)\det ({\bf 1}+\bar{Z}Z)^{-2K-N}
\eea
being the resolution of unity in the so defined Hilbert space. Note that this represents a compact phase space integration, the case $N=1$, $K=1$ being the stereographic sphere. The trace now follows
\bea
\Tr_{K,N}(\hat{O})&=&\int[d\bar{Z}dZ]_{K,N}\langle\bar{Z}|\hat{O}|Z\rangle
\eea
We will use this trace to consider evaluation of the partition function
\bea
Z(\beta)&=&\Tr\left(e^{-\beta \hat{H}}\right)
\eea
for the simple case where our Hamiltonian belongs to the group algebra
\bea
\hat{H}=\omega\sum_k \hat{B}(k,k)\equiv\omega\hat{B}.
\eea
\RD{The Euclidean thermal circle enters only through the operator trace $\Tr(e^{-\beta \hat H})$. It is not included in $K$, which continues to count only the regulated spatial modes.} From the commutation relations
\bea
\left[\hat{A}^\dagger(k,l),\hat{B}\right]&=&A^\dagger(k,l)
\eea
we have
\bea
e^{-\omega\beta\hat{B}}|Z\rangle &=& |e^{-\omega\beta}Z\rangle
\eea
and
\bea
Z(\beta)&=&\int [dZd\bar{Z}]_{K,N}\langle\bar{Z}|e^{-\beta\omega}Z\rangle
\eea
or
\bea
Z(\beta)&=&C(N,K)\int (dZd\bar{Z})\det(1+Z\bar{Z})^{-2K-N}\det(1+x\bar{Z}Z)^N
\eea
where $x=e^{-\beta\omega}$. Following the diagonalization as in~\cite{Das:2012dt,Ginibre:1965zz} we have
\bea
Z(\beta)&=&C(N,K){{\rm Vol}\,\Omega\over K!}\int \prod_{l=1}^K d\omega_l \,\,\Delta^2(\omega_1^2,\omega_2^2,\cdots,\omega_K^2)\,\prod_l(1+\omega_l^2)^{-2K-N}\cr\cr
&&\qquad\qquad\qquad\qquad\qquad\qquad\times\prod_l(1+x\omega_l^2)^N
\eea
with normalization $C(N,K)$ being
\bea
C(N,K)^{-1}&=&{{\rm Vol}\,\Omega\over K!}\int \Delta^2(\omega_1^2,\omega_2^2,\cdots, \omega_K^2)\prod_l(1+\omega_l^2)^{-2K-N}\,,
\eea
${\rm Vol}\,\Omega$ is the volume of the ``angular'' parts of the integration and $\Delta(x_1,\cdots,x_K)=\prod_{k<l}(x_k-x_l)$ is the Vandermonde determinant, with $x_i=\omega_i^2$. Changing variables
\bea
\omega_i^2&=&\frac{y_i}{1-y_i}
\eea
we have the final $K$-integral expression for the partition function
\bea
Z(\beta)&=&\frac{{\rm Vol}\,\Omega}{2^KK!}C(N,K)\int_0^1\prod_{l=1}^K dy_i\Delta^2(y_1,y_2, \cdots,y_K)\prod_{i=1}^K (1-(1-x)y_i)^N
\eea
Evaluating this partition function amounts to evaluating the $K$--fold integral
\begin{equation}
I_{K,N}(x)\;=\;\int_{0}^{1}\cdots\int_{0}^{1}\Bigl(\Delta(y_1,\dots,y_K)\Bigr)^{2}
\prod_{i=1}^{K}(1-(1-x)y_i)^N\,dy_i.
\end{equation}
We will make use of two identities: the explicit expression for $\Delta(y_1,\dots,y_K)$ as a determinant
\begin{eqnarray}
\Delta(y_1,\dots,y_K)&=&\det\bigl[y_i^{\,j-1}\bigr]_{i,j=1}^{K}
\end{eqnarray}
and Andr\'eief's integration formula~\cite{Forrester}, which says
\begin{eqnarray}
&&\int_{I}\cdots\int_{I}\det\bigl[f_{j-1}(t_i)\bigr]_{i,j=1}^{K}\,\det\bigl[\phi_{j-1}(t_i)\bigr]_{i,j=1}^{K}\,dt_1 \cdots dt_K\cr\cr
&&\qquad\qquad=\,K!\,\det\Bigl[\int_I f_i(t)\phi_j(t)\,dt\Bigr]_{i,j=0}^{K-1}.
\end{eqnarray}
where $f_0,\dots,f_{K-1}$ and $\phi_0,\dots,\phi_{K-1}$ are integrable functions on the interval $I$. This turns the $K$--fold integral we need to evaluate into a single determinant of \emph{one--fold} integrals. In terms of the weight
\begin{equation}
w(y)=(1-(1-x)y)^N\,.
\end{equation}
we can write the integrand as
\begin{equation}
\Delta(y)^2 \prod_{i=1}^K w(y_i)=\det[y_i^{j-1}]\,\det[y_i^{j-1}]\;\prod_{i=1}^K w(y_i).
\end{equation}
Now absorb $w(y_i)$ into the first determinant by setting
\begin{equation}
f_{j-1}(y)=y^{j-1}w(y),\qquad \phi_{j-1}(y)=y^{j-1}, \qquad (j=1,\dots,K).
\end{equation}
Then Andr\'eief's formula gives
\begin{equation}
I_{K,N}(x)=K!\,\det\!\Bigl[\mu_{i+j}\Bigr]_{i,j=0}^{K-1},\qquad
\mu_m=\int_{0}^{1} y^{m}\,w(y)\,dy=\int_0^1 y^m(1-(1-x)y_i)^N\,dy.
\end{equation}
This determinant is a \emph{Hankel determinant}, i.e. its the determinant of a matrix whose entries depend only on the sum of the row and column indices (above entries depend only on $i+j$). So the problem is now to compute the moments $\mu_m$. Recalling Euler's beta function
\begin{equation}
B(a,b)=\int_0^1 t^{a-1}(1-t)^{b-1} dt=\frac{\Gamma(a)\Gamma(b)}{\Gamma(a+b)}.
\end{equation}
and the Euler integral representation of the Gauss hypergeometric function: for ${\rm Re}(C)>{\rm Re}(B)>0$,
\begin{equation}
\,{}_2F_1(A,B;C;\xi)=\frac{\Gamma(C)}{\Gamma(B)\Gamma(C-B)}\int_0^1 t^{B-1}(1-t)^{C-B-1}(1-\xi t)^{-A} dt.
\end{equation}
Match this to our moment integral by taking
\begin{equation}
A=-N,\qquad B=m+1,\qquad C=m+2.
\end{equation}
Then the integral representation becomes
\begin{equation}
\int_0^1 t^{m}(1-(1-x)t)^{-N}dt=\frac{1}{m+1}\,{}_2F_1\!\bigl(-N,m+1;m+2;1-x\bigr)
\end{equation}
and we obtain
\begin{equation}
\mu_m=\frac{1}{m+1}\,\,{}_2F_1\bigl(-N,m+1;m+2;1-x\bigr).
\end{equation}
Thus we can write the following closed form for our partition function
\begin{equation}
Z(\beta)=N_Z\,\det \Big[\frac{1}{i+j+1}\,{}_2F_1\bigl(-N,i+j+1;i+j+2;1-x\bigr)\Big]_{i,j=0}^{K-1}.
\end{equation}
with $N_Z$ a constant given by
\bea
N_Z={\rm Vol}\,\Omega\,\,C(N,K)
\eea
We plot this partition function below. For a finite-dimensional Hilbert space, the qualitative behaviour in Fig.~\ref{fig:Z} is precisely what one expects for the logarithm of the canonical partition function. Writing $Z(T)=\Tr\, e^{-H/T}$, the Boltzmann suppression of excited states weakens as $T$ increases, so additional levels contribute and $Z(T)$ grows, yielding a monotonically increasing curve. Moreover, finiteness of the spectrum implies $Z(T)\to \dim\mathcal H(N,K)$ as $T\to\infty$, and hence $\log Z(T)$ saturates. The observed concave-down approach to a plateau at large temperature is the corresponding manifestation of this high-$T$ limit.

\bigskip

\section{Conclusions}\label{Conclusions}

We have discussed in this work two descriptions of the Hilbert space for finite $N$ singlets in vector theories: (i) the bilocal formulation, in which the space of configurations is viewed as a K\"ahler phase space and quantized using geometric quantization, and (ii) the direct construction in terms of gauge-invariant operators, whose counting is captured by the Molien--Weyl formula. We showed that these perspectives are equivalent, in the sense that the number of independent invariants coincides with the dimension of the Hilbert space\footnote{The proposal for the Hilbert space made in~\cite{Anninos:2017eib} does not agree with the finite $N$ Hilbert space of either bosons or fermions. In particular, that proposal misses the secondary class of states.}, obtained through (overcomplete) geometric\cite{Das:2012dt} quantization. Finally a third specification of the Hilbert space is introduced, based on group of invariants; the details of this will be presented in the companion  publication~\cite {Us2}.

\begin{center} 
{\bf Acknowledgements}
\end{center}
We would like to acknowledge discussions with Sumit Das and Junjie Zheng. We are  grateful to Junjie Zheng for correcting an error in the earlier version. The work of RdMK was supported by a start up research fund of Huzhou University, a Zhejiang Province talent award and by a Changjiang Scholar award. J.Y. was supported by the National Research Foundation of Korea (NRF) grant funded by the Korean government (MSIT) (RS-2022-NR069038) and by the Brain Pool program funded by the Ministry of Science and ICT through the National Research Foundation of Korea (RS-2023-00261799).

\appendix
\section{Other groups}\label{intevaluate}

In this appendix we extend the same analysis to the gauge groups $SO(N)$ and $Sp(N)$.

\subsection{\texorpdfstring{$Sp(N)$}{Sp(N)} gauge group}

Consider the count of invariants, performed using the Molien-Weyl formula. We have a model of $K$ fermionic fields in the fundamental representation of $Sp(N)$. Using standard Molien-Weyl formulas, the exact Hilbert series is
\bea
Z(x)&=&{1\over N! 2^N(2\pi i)^N}\oint \prod_{j=1}^N {d\varepsilon_{+j}\over\varepsilon_{+j}}\,\,\Delta_+\Delta_- 
\prod_{i=1}^N\, (1+\varepsilon_{+i}x)^K(1+\varepsilon_{+i}^{-1}x)^K\label{finalPF}
\eea
where
\bea
\Delta_+&=&\prod_{1\le k<r\le N}(\varepsilon_{+k}-\varepsilon_{+r})(1-\varepsilon_{+k}\varepsilon_{+r})\prod_{j=1}^N (\varepsilon_{+j}-\varepsilon_{+j}^{-1})
\eea
\bea
\Delta_-&=&\prod_{1\le k<r\le N}(\varepsilon_{+k}^{-1}-\varepsilon_{+r}^{-1})(1-\varepsilon_{+k}^{-1}\varepsilon_{+r}^{-1})\prod_{j=1}^N (\varepsilon_{+j}^{-1}-\varepsilon_{+j})
\eea
Each integral over the variables $\epsilon_{+j}$ is performed along the unit circle in the complex plane and can be evaluated using the residue theorem. We can simplify the above result for $Z(x)$ as follows: Change to the angle $\theta_j\in[0,2\pi)$ defined by $\varepsilon_{+j}=e^{i\theta_j}$ to obtain 
\bea
Z(x)&=&{1\over N! 2^N}\int_{[0,2\pi]^N}\,\, \prod_{q=1}^N {d\theta_q\over 2\pi}
\prod_{1\le i<j\le N} |e^{i\theta_i}-e^{i\theta_j}|^2\,|1-e^{i(\theta_i+\theta_j)}|^2\,
\prod_{k=1}^N |1-e^{2i\theta_k}|^2\cr\cr
&&\qquad\times \prod_{l=1}^N (1+2x\cos\theta_l+x^2)^K
\eea
The integrand is even and $\pi$-periodic in each $\theta_j$ so we can restrict to $\theta_j\in[0,\pi]$
\bea
Z(x)&=&{1\over N!}\int_{[0,\pi]^N}\,\, \prod_{q=1}^N {d\theta_q\over 2\pi}
\prod_{1\le i<j\le N} |e^{i\theta_i}-e^{i\theta_j}|^2\,|1-e^{i(\theta_i+\theta_j)}|^2\,
\prod_{k=1}^N |1-e^{2i\theta_k}|^2\cr\cr
&&\qquad\times \prod_{l=1}^N (1+2x\cos\theta_l+x^2)^K
\eea
Now make a further change of variables to $t_j$ defined by
\bea
t_j=\sin^2\!\frac{\theta_j}{2}\ \in[0,1],\qquad\cos\theta_j=1-2t_j,\qquad d\theta_j=\frac{dt_j}{\sqrt{t_j(1-t_j)}}
\eea
to obtain
\bea
Z(x)&=&{2^{2N(N+1)} (1+x)^{2NK}\over N! (2\pi)^N}\int_{[0,1]^N}\,\, \prod_{q=1}^N d t_q \sqrt{t_q(1-t_q)}\left(1-\frac{4xt_q}{(1+x)^2}\right)^K \Delta(t)^2\cr
&&
\eea
where
\bea
\Delta(t)&\equiv&\prod_{1\le i<j\le N} (t_i-t_j)
\eea
To compute the total number of invariants we evaluate this expression at $x=1$ to find
\bea
Z(1)&=&{2^{2N(N+1)} 2^{2NK}\over N! (2\pi)^N}\int_{[0,1]^N}\,\, \prod_{q=1}^N d t_q t_q^{1\over 2}(1-t_q)^{K+{1\over 2}}\Delta(t)^2
\eea
Comparing to the Selberg integral
\bea
I(\alpha,\beta,\gamma,n)&=&\int_0^1 dx_1\cdots \int_0^1dx_n \vert\Delta(x)\vert^{2\gamma} \prod_{j=1}^n x_j^{\alpha-1}(1-x_j)^{\beta-1}\cr
&=&\prod_{j=0}^{n-1}\frac{\Gamma(1+\gamma+j\gamma)\Gamma(\alpha+j\gamma)\Gamma(\beta+j\gamma)}{\Gamma(1+\gamma)\Gamma(\alpha+\beta+(n+j-1)\gamma)}
\eea
we easily see that
\bea
Z(1)&=&{2^{2N(N+1)} 2^{2NK}\over N! (2\pi)^N}I({3\over 2},K+{3\over 2},1,N)\cr\cr
&=&{2^{2N(N+1)} 2^{2NK}\over N! (2\pi)^N}
\prod_{j=0}^{N-1}\frac{\Gamma(2+j)\Gamma({3\over 2}+j)\Gamma(K+{3\over 2}+j)}{\Gamma(2)\Gamma(K+2+N+j)}
\eea
Now use the $\Gamma$ function identity
\bea
\Gamma(z+{1\over 2})&=&{2^{1-2z}\sqrt{\pi}\Gamma(2z)\over\Gamma(z)}
\eea 
to obtain
\bea
Z(1)&=&{2^N\over N!}\prod_{j=0}^{N-1}{\Gamma(2+j)\Gamma(2(j+1))\Gamma(2(k+1+j))\over \Gamma(k+N+2+j)\Gamma(j+1)\Gamma(K+1+j)}\label{finalcount}
\eea

We would now like to match this counting to the dimension of a Hilbert space obtained by K\"ahler quantization of a suitable phase space. For $Sp(2N)$ gauge symmetry, the basic bilocal invariant is symmetric in site indices (two Grassmann minus signs cancel the antisymmetry of the symplectic form), so we should uses a \emph{complex symmetric} matrix
\bea
  Z^T&=&Z\,,\qquad
  dZ\,d\bar Z \equiv \prod_{1\le p\le q\le K} d^2 Z_{pq}\,.
\eea
The coherent-state overlap takes the determinant form
\bea
  \langle Z|W\rangle &=&\det\,\bigl({\bf 1}+Z^\dagger W\bigr)^N\,,
  \qquad\langle Z|Z\rangle\,\,=\,\,\det\,\bigl({\bf 1}+Z^\dagger Z\bigr)^N\,,
\eea
so the quantization level is $\lambda=N$. The Berezin measure that matches the $Sp(2N)$ Molien--Weyl counting is then
\bea
d\mu(Z,\bar Z)&=& \frac{1}{C(N,K)}\,\prod_{1\le p\le q\le K} d^2 Z_{pq}\,\det\,\bigl({\bf 1}+Z Z^\dagger\bigr)^{-N-(K+1)}\,.
\label{eq:berezin-measure-Sp}
\eea
For the dimension of the Hilbert space produced with this quantization, insert \eqref{eq:berezin-measure-Sp} into $\dim\mathcal H=\int d\mu\,\langle Z|Z\rangle$ and reduce the resulting integral to eigenvalues using \cite{Takagi}
\begin{equation}
  Z \;=\; U\,\mathrm{diag}(r_1,\dots,r_K)\,U^T,\qquad U\in U(K),\qquad r_i\ge 0,
\end{equation}
followed by the standard change of variables $t_i=\frac{r_i^2}{1+r_i^2}\in[0,1]$. After integrating out the angular $U(K)$, the remaining $t$-integral is precisely a Selberg integral with the same parameters as in the Molien--Weyl evaluation. Concretely, we find
\bea
\dim\mathcal H &\propto& \int_{[0,1]^N}\prod_{q=1}^N dt_q\, t_q^{\frac12}(1-t_q)^{K+\frac12}\,\Delta(t)^2\,\,=\,\, I\,\left(\frac32,K+\frac32,1,N\right),
\eea
Thus the K\"ahler/Berezin trace computation with \eqref{eq:berezin-measure-Sp}
reduces to the same Selberg integral and thus reproduces \eqref{finalcount}.

\subsection{\texorpdfstring{$SO(N)$}{SO(N)} gauge group}

Consider $K$ fermionic fields $\psi_p^i$ ($p=1,\dots,K$, $i=1,\dots,N$) transforming in the defining (vector) representation of $O(N)$. The full Fock space is
\bea
\mathcal H &\cong& \bigwedge\nolimits^\bullet\!\big(\mathbb C^N\otimes \mathbb C^K\big).
\eea
Gauge-invariants are obtained by contracting gauge indices with $\delta_{ij}$. Introduce the graded counting function
\bea
Z(x)&=&\sum_{F\ge 0} x^F\,\dim\big(\mathcal H_F^{\,O(N)}\big)
\eea
where $\mathcal H_F$ is the $F$-fermion sector. Using the identity
\bea
\sum_{n\ge 0} x^n\,\chi_{\wedge^n \mathbf N}(U)&=&\det({\bf 1}+xU),
\eea
the Molien--Weyl formula gives
\bea
Z(x)&=&\int_{O(N)} dU\,\det({\bf 1}+xU)^{\,K}.
\eea
There is a useful simplification at $x=1$. For $K>0$, every matrix in the disconnected component $\det(U)=-1$ has at least one eigenvalue $-1$, hence $\det({\bf 1}+U)=0$ on that component. As a result,
\bea
Z_{O(N)}(1)&=&\frac12\,Z_{SO(N)}(1)\qquad (K>0).
\eea
For completeness: if $K=0$, then $Z(1)=1$ trivially, and the ``$\tfrac12$'' should be replaced by $1$. It remains to evaluate $Z_{SO(N)}(1)$, which depends on whether $N$ is even or odd.

\subsubsection{\texorpdfstring{$O(2N)$}{O(2N)}}

Start from the Weyl integration formula for $SO(2N)$ written in angular variables $\theta_j\in[0,\pi]$. With
\bea
\det({\bf 1}+U)=\prod_{j=1}^N\big(2+2\cos\theta_j\big)&=&4^N\prod_{j=1}^N \cos^2\!\frac{\theta_j}{2},
\eea
set $t_j=\sin^2(\theta_j/2)\in[0,1]$. We finds a Selberg integral with parameters
\bea
\alpha&=&\frac12,\qquad \beta\,\,=\,\,K+\frac12,\qquad \gamma\,\,=\,\,1,
\eea
so that (for $K>0$)
\bea
Z_{O(2N)}(1)&=&\frac{1}{2}\,Z_{SO(2N)}(1)\cr\cr
&=&\frac{2^{(N-1)(2N-1)+2NK-1}}{\pi^N\,N!}\, I\left(\tfrac12,\,K+\tfrac12,\,1,\,N\right),
\eea
where $I(\alpha,\beta,\gamma,N)$ is the Selberg integral in the notation introduced above. Using Selberg's closed product and simplifying the half-integer $\Gamma$-functions, the final answer can be written as the factorial product
\bea
Z_{O(2N)}(1)&=&\frac{1}{2^{N}\,N!}\,\prod_{j=0}^{N-1}\frac{(j+1)!\,(2j)!\,(2K+2j)!}{j!\,(K+j)!\,(K+N+j-1)!}
\qquad (K\ge 1)\label{SOPFnctn}
\eea

\subsubsection{\texorpdfstring{$O(2N+1)$}{O(2N+1)}}

For $SO(2N+1)$ there is an additional fixed eigenvalue $+1$, so
\bea
\det({\bf 1}+U)&=&(1+1)\prod_{j=1}^N (2+2\cos\theta_j)\,\,=\,\,2\cdot 4^N\prod_{j=1}^N \cos^2\!\frac{\theta_j}{2}.
\eea
The Weyl formula now includes an extra $\prod_j \sin^2(\theta_j/2)$ factor. With the same change of variables $t_j=\sin^2(\theta_j/2)$, we get a Selberg integral with
\bea
\alpha&=&\frac32,\qquad \beta\,\,=\,\,K+\frac12,\qquad \gamma\,\,=\,\,1,
\eea
and hence (for $K>0$)
\bea
Z_{O(2N+1)}(1)&=&\frac{1}{2}\,Z_{SO(2N+1)}(1)\cr\cr
&=&\frac{2^{N(2N-1)+(2N+1)K-1}}{\pi^N\,N!}\,I\left(\tfrac32,\,K+\tfrac12,\,1,\,N\right).
\eea
Selberg evaluation simplifies to the factorial product
\bea
Z_{O(2N+1)}(1)&=&\frac{2^{K-N-1}}{N!}\,\prod_{j=0}^{N-1}\frac{(2j+2)!\,(2K+2j)!}{(K+j)!\,(K+N+j)!}\qquad (K\ge 1).
\eea
Note that for $O(2)$ ($N=1$) the even-$N$ result gives
\bea
Z_{O(2)}(1)&=&\frac12\binom{2K}{K}.
\eea
For $O(1)$ the odd-$N$ formula with $N=0$ gives $Z_{O(1)}(1)=2^{K-1}$, i.e.\ only even-fermion states survive the $\mathbb Z_2$ gauge projection.

Finally, we again want to see how this count of invariants matches the dimension of a Hilbert space obtain by K\"ahler quantization. For $O(N)$ gauge symmetry, contraction uses the symmetric tensor $\delta_{ij}$, so the fermionic bilocal invariant is \emph{antisymmetric} in site indices. Thus one uses a \emph{complex antisymmetric} matrix
\bea
  Z^T&=&-Z\,,\qquad dZ\,d\bar Z \equiv \prod_{1\le p<q\le K} d^2 Z_{pq}\,.
\eea
For fermionic pairing coherent states, the overlap carries a square
root of a determinant for $N=1$. With $N$ colors this gives
\bea
\langle Z|W\rangle &=&\det\,\bigl({\bf 1}+Z^\dagger W\bigr)^{\frac{N}{2}},\qquad
\langle Z|Z\rangle\,\,=\,\,\det\,\bigl({\bf 1}+Z^\dagger Z\bigr)^{\frac{N}{2}}.
\eea
So the quantization level is $\lambda=\frac{N}{2}$. For odd $N$ this is
half-integer and is exactly what is needed to reproduce the even/odd $O(N)$
distinction in the Molien--Weyl result. The Berezin measure that matches the $O(N)$ Molien--Weyl counting is
\bea
d\mu_{O}(Z,\bar Z)&=&\frac{1}{C(N,K)}\,\prod_{1\le p<q\le K} d^2 Z_{pq}
\det\,\bigl({\bf 1}+Z Z^\dagger\bigr)^{-\frac{N}{2}-(K-1)}\,.
\label{eq:berezin-measure-O}
\eea
Repeat the trace computation $\dim\mathcal H=\int d\mu\,\langle Z|Z\rangle$ with
\eqref{eq:berezin-measure-O}. Reduce the integral to eigenvalues using \cite{Youla}
\bea
Z&=&U\,\bigoplus_{a=1}^{\lfloor K/2\rfloor}
  \begin{pmatrix}
    0 & r_a\\
    -r_a & 0
  \end{pmatrix}
  \,U^T,
  \qquad U\in U(K),\qquad r_a\ge 0,
\eea
and again set $t_a=\frac{r_a^2}{1+r_a^2}\in[0,1]$.
After integrating out the angular $U(K)$, the remaining eigenvalue integral is
a Selberg integral which differs for $O(2N)$ vs $O(2N{+}1)$
\begin{align}
  O(2N):\,\,&\dim\mathcal H\;\propto\;\int_{[0,1]^N}\prod_{q=1}^N dt_q\;
  t_q^{-\frac12}(1-t_q)^{K-\frac12}\,\Delta(t)^2\;=\;
  I\!\left(\frac12,K+\frac12,1,N\right),
  \\
  O(2N{+}1):\,\,&\dim\mathcal H\;\propto\;\int_{[0,1]^N}\prod_{q=1}^N dt_q\;
  t_q^{+\frac12}(1-t_q)^{K-\frac12}\,\Delta(t)^2\;=\;
  I\!\left(\frac32,K+\frac12,1,N\right).
\end{align}
These are exactly the Selberg integrals produced by the corresponding Molien--Weyl integrals at $x=1$ for $O(2N)$ and $O(2N{+}1)$, so the K\"ahler-quantization Hilbert-space dimension again agrees with the invariant
count (as in the $U(N)$ check of Section \ref{HilbertSpace}).

\section{Equivalence of trace relations and the rank constraint}\label{EquivRels}

In this Appendix we discuss a simple example that explicitly tests the equivalence of the standard description of the trace relations, employing Procesi's~\cite{Pr} universal multilinear identity, and the rank constraint \eqref{flavorrelation}. As a concrete example, take $N=2$ and $K=3$. A standard count of color invariants using the Molien--Weyl formula (derived for bosonic vector models in~\cite{deMelloKoch:2025cec}) gives
\bea
H(x)&=&\frac{1+x+x^2}{(1-x)^8}\label{HbyProcesi}
\eea
This gives a count of all invariants after Procesi's universal multilinear identity has been imposed.

It is straightforward to reproduce the same result directly from (\ref{flavorrelation}). For $K=3$, $\alpha(k,l)$ is a $3\times 3$ matrix and (\ref{flavorrelation}) becomes
\bea
\det(\alpha)&=&0\,.
\eea
This imposes a single cubic relation on the ring generated by the $9$ entries of $\alpha(k,l)$, so the Hilbert series is
\bea
H(x)&=&\frac{1-x^3}{(1-x)^9}\,.
\eea
This exactly reproduces (\ref{HbyProcesi}) once a factor of $(1-t)$ is cancelled between numerator and denominator. A completely parallel exercise can be repeated for other values of $K$ and $N$, and also for the fermionic models.

\section{Quotienting by the trace relations using Macaulay2}\label{Macaulay2}

Macaulay2 is an open source software system, hosted on the Melbourne Research Cloud, at this URL:
\begin{center}
https://www.unimelb-macaulay2.cloud.edu.au/\#home
\end{center}
Macaulay2 supports research in algebraic geometry, commutative algebra, and related fields in mathematics or applications. Using Macaulay2 we can specify the free polynomial ring $R$ generated by the bilinears and the ideal $I$ generated by the relations (R1) and (R2), and then analyze the quotient algebra $A=R/I$. Our claim of Section \ref{BSection3} is that the Hilbert space generated by the collective bilinear is naturally identified with this quotient. We can test this claim directly in Macaulay2 by constructing $A$, computing its Hilbert series, and extracting explicit graded bases. These computations verify that the graded dimensions and basis elements of $A$ match the expected state counting, confirming that $A$ indeed realizes the Hilbert space generated by the collective bilinear.

\paragraph{Results for $p=4$:}
We run the following script from the web interface for Macaulay2:
\begin{verbatim}
R = QQ[z12,z13,z14,z23,z24,z34, Degrees => {1,1,1,1,1,1}];
I = ideal(
  z12^2, z13^2, z14^2, z23^2, z24^2, z34^2, z12*z13,
  z12*z14, z13*z14, z12*z23, z12*z24, z23*z24, z13*z23, 
  z13*z34, z23*z34, z14*z24, z14*z34, z24*z34, 
  z12*z34 - z14*z23, z13*z24 + z14*z23);
A = R/I;
hilbertSeries A
for d from 0 to 6 list (d, hilbertFunction(d,A))
sum for d from 0 to 10 list hilbertFunction(d,A)
basis(0,A)
basis(1,A)
basis(2,A)
basis(3,A)
\end{verbatim}
The first thing the script does is to set up a commutative algebra $R$ in the variables $z_{12}$, $z_{13}$, $z_{14}$, $z_{23}$, $z_{24}$, $z_{34}$. The code then defines the ideal $I$ (comprising of the relations that vanish). $I$ is the set of relations
\begin{eqnarray}
I&=&\{ z_{12}^2, z_{13}^2, z_{14}^2, z_{23}^2, z_{24}^2, z_{34}^2, z_{12}z_{13}, z_{12}z_{14},
  z_{13}z_{14}, z_{12}z_{23}, z_{12}z_{24}, z_{23}z_{24}, z_{13}z_{23},   \cr\cr    
 && z_{13}z_{34}, z_{23}z_{34}, z_{14}z_{24}, z_{14}z_{34}, z_{24}z_{34}, z_{12}z_{34} - z_{14}z_{23}, z_{13}z_{24} + z_{14}z_{23}\}
\end{eqnarray}
The code produces the Hilbert series
\begin{equation}
H=1+6T+T^2
\end{equation}
indicating that the Hilbert space is 8 dimensional with 1 degree zero state, 6 degree 2 states and 1 degree 2 state. The bilinear is assigned degree 1. The script also produces the following explicit basis
\begin{equation}
\{1,z_{12},z_{13},z_{14},z_{23},z_{24},z_{34},z_{12}z_{34}\}
\end{equation}
This exactly reproduces the bilocal Hilbert space.

\paragraph{Results for $p=6$:}
We run the following script:
\begin{verbatim}
R = QQ[z12,z13,z14,z15,z16,z23,z24,z25,z26,z34,z35,z36,z45,z46,z56,
  Degrees => {1,1,1,1,1, 1,1,1,1, 1,1,1, 1,1, 1}];
I = ideal(  
z12^2, z13^2, z14^2, z15^2, z16^2, z23^2, z24^2, z25^2, z26^2, 
z34^2, z35^2, z36^2, z45^2, z46^2, z56^2, z12*z13, z12*z14, 
z12*z15, z12*z16,  z13*z14, z13*z15, z13*z16, z14*z15, z14*z16, 
z15*z16, z12*z23, z12*z24, z12*z25, z12*z26,  z23*z24, z23*z25,
z23*z26, z24*z25, z24*z26, z25*z26, z13*z23, z13*z34, z13*z35,
z13*z36,  z23*z34, z23*z35, z23*z36, z34*z35, z34*z36, z35*z36, 
z14*z24, z14*z34, z14*z45, z14*z46,  z24*z34, z24*z45, z24*z46, 
z34*z45, z34*z46, z45*z46, z15*z25, z15*z35, z15*z45, z15*z56, 
z25*z35, z25*z45, z25*z56, z35*z45, z35*z56, z45*z56, z16*z26, 
z16*z36, z16*z46, z16*z56, z26*z36, z26*z46, z26*z56, z36*z46, 
z36*z56, z46*z56, z12*z34 - z14*z23, z13*z24 + z14*z23,  
z12*z35 - z15*z23, z13*z25 + z15*z23,  z12*z36 - z16*z23, 
z13*z26 + z16*z23,  z12*z45 - z15*z24, z14*z25 + z15*z24,  
z12*z46 - z16*z24, z14*z26 + z16*z24, z12*z56 - z16*z25,  
z15*z26 + z16*z25,  z13*z45 - z15*z34, z14*z35 + z15*z34,  
z13*z46 - z16*z34, z14*z36 + z16*z34, z13*z56 - z16*z35,  
z15*z36 + z16*z35, z14*z56 - z16*z45, z15*z46 + z16*z45, 
z23*z45 - z25*z34, z24*z35 + z25*z34, z23*z46 - z26*z34,  
z24*z36 + z26*z34, z23*z56 - z26*z35, z25*z36 + z26*z35, 
z24*z56 - z26*z45,  z25*z46 + z26*z45,z34*z56 - z36*z45,  
z35*z46 + z36*z45);
A = R/I;
hilbertSeries A
for d from 0 to 6 list (d, hilbertFunction(d, A))
sum for d from 0 to 10 list hilbertFunction(d, A)
for d from 0 to 3 list (d, numColumns basis(d, A))
basis(0,A)
basis(1,A)
basis(2,A)
basis(3,A)
basis(4,A)
basis(5,A)
\end{verbatim}
Now $R$ is a free commutative algebra with 15 generators and the ideal $I$ has 105 relations. The code produces the Hilbert series
\begin{equation}
H=1+15T+15T^2+T^3
\end{equation}
and the following explicit basis (separated by degree for readability)
\begin{eqnarray}
&&\!\!\!\!\!\!\!\!\!\{1,\cr
&&\!\!\!\!\!\!z_{12},z_{13},z_{14},z_{15},z_{16},z_{23},z_{24},z_{25},z_{26},z_{34},z_{35},z_{36},z_{45},z_{46},z_{56},\cr
&&\!\!\!\!\!\!z_{12}z_{34},z_{12}z_{35},z_{12}z_{36},z_{12}z_{45},z_{12}z_{46},z_{12}z_{56},z_{13}z_{45},z_{13}z_{46},z_{13}z_{56},\cr
&&\!\!\!\!\!\!z_{14}z_{56},z_{23}z_{45},z_{23}z_{46},z_{23}z_{56},z_{24}z_{56},z_{34}z_{56},\cr
&&\!\!\!\!\!\!z_{12}z_{34}z_{56}\}
\end{eqnarray}
which again exactly reproduces the bilocal Hilbert space.

\end{document}